# Kilometre-scale Jovian moon characterised for a potential JUICE flyby

Juan Luis Rizos[1*], Jose María Gómez-Limón[1], Yucel Kilic[1,2], Altair R. Gomes-Júnior[3,4], Jose Luis Ortiz[1], Nicolas Morales[1], Álvaro Álvarez-Candal[1], Pasquale Palumbo[5], Luisa M. Lara[1], Simon Porter[6], Benjamin Proudfoot[7], Evelyn Christiny M. Prais[3], Jean Marc Christille[8], Stefano Sartor[8], Antonio Cabrera[9], Stefan Geier[9], Alessia Spolon[10], Alice Lucchetti[10], Andrea Soffiantini[11], Anne Verbiscer[6], Armin Behrend[12], Elena Mazzotta Epifani[13], Jean-Louis Dumont[14], Luca Zampieri[10], Mike Skrutskie[15], Maurizio Pajola[10], Michel Boutet[16], Michele Fiori[10], Raoul Behrend[17], Stefano Sposetti[18], Vadim Nikitin[19], Fabien Cavaille[20], Jocelyn Sérot[21], Laurent Miralabe[22], Marc Buie[23], Mehmed Naim Bagiran[24], Metin Altan[25], Mohammad Niaei[26], Noël Robichon[27], Orhan Erece[28,29], Pierre-Louis Phan[27], Raphael Lallemand[27], Süleyman Fişek[30], Valentin Etienne[27], Valery Lainey[27], Yavuz Güney[31], Yusuf Avcioglu[30], Francois Poulet[32], Paul Hartogh[33], Oliver Witasse[34], Pablo Santos-Sanz[1], Julia de León[35,36], Flavia L. Rommel[7], Fernando Tinaut-Ruano[37]

[1]Instituto de Astrofísica de Andalucía, Solar System Department, CSIC, Glorieta de la Astronomía s/n, E-18008 Granada, Spain.
[2]LIRA, CNRS UMR 8254, Observatoire de Paris, Meudon, France.
[3]Institute of Physics, Federal University of Uberlândia, Av. João Naves de Ávila, Uberlândia, 38408-100, MG, Brazil.
[4]Laboratório Interinstitucional de e-Astronomia - LIneA, Av. Pastor Martin Luther King Jr, 126, Rio de Janeiro, 20765-000, RJ, Brazil.
[5]INAF-IAPS, Istituto di Astrofisica e Planetologia Spaziali di Roma, Rome, Italy.
[6]Southwest Research Institute, 1301 Walnut Street, Suite 400, Boulder, CO 80302, USA.

[7]Florida Space Institute, University of Central Florida, 12354 Research Parkway, Partnership I, Room 211, 32826 Orlando, USA.
[8]Astronomical Observatory of the Autonomous Region of Aosta Valley (OAVdA), Fondazione Clément Fillietroz, Nus, Italy.
[9]GRANTECAN, Cuesta de San José s/n, E-38712, Breña Baja, La Palma, Spain.
[10]INAF-Astronomical Observatory of Padova, Padova, Italy.
[11]Unione Astrofili Bresciani - Osservatorio Astronomico Serafino Zani di Lumezzane, Brescia, Italy.
[12]Miam-Globs Observatory, Les Verrières, Switzerland.
[13]INAF-Osservatorio Astronomico di Roma, Rome, Italy.
[14]Société Astronomique de Touraine, Tauxigny, France.
[15]Department of Astronomy, University of Virginia, P.O. Box 400325, Charlottesville, VA 22904-4325, USA.
[16]Ploumilliau Observatory, Ploumilliau, France.
[17]Observatoire de Genève, Geneva, Switzerland.
[18]IOTA/ES, Europe.
[19]IOTA-US, USA.
[20]Observatory of Sassay, Sassay, France.
[21]Observatory of Petit Cosquet, France.
[22]Société d'Astronomie de Rennes, Rennes, France.
[23]Southwest Research Institute, 1050 Walnut St 300, Boulder, CO, 80302, USA.
[24]TÜRKSAT Satellite Communication Cable TV and Management Inc., Yağlıpınar Dist. Gölbası, Ankara, Türkiye.
[25]Eskisehir Technical University, Astrophysics Education and Research Unit, Yunusemre Observatory, Eskişehir, Türkiye.
[26]Ankara University, Science Faculty, Department of Astronomy and Space Sciences, 06100 Ankara, Türkiye.
[27]Observatoire de Paris / Paris Observatory, Paris, France.
[28]Türkiye National Observatories, TUG, 07070, Antalya, Türkiye.
[29]The Scientific and Technological Research Council of Türkiye (TÜBİTAK), 06680 Ankara, Türkiye.
[30]Department of Astronomy and Space Sciences, Faculty of Science, Istanbul University, 34116 Istanbul, Türkiye.
[31]Ataturk University, Science Faculty, Department of Physics, 25240 Erzurum, Türkiye.
[32]Institut d'Astrophysique Spatiale, CNRS/Paris-Saclay University, France.

[33]Max Planck Institute for Solar System Research, Justus-von-Liebig-Weg 3, 37077, Göttingen, Germany.
[34]European Space Research and Technology Centre, European Space Agency, 2201 AZ, Noordwijk, The Netherlands.
[35]Instituto de Astrofísica de Canarias, C/Vía Láctea s/n, La Laguna, 38205, Tenerife, Spain.
[36]Departamento de Astrofísica, Universidad de La Laguna, Avda. Astrofísico Francisco Sánchez s/n, La Laguna, 38200, Tenerife, Spain.
[37]Université Côte d'Azur, Observatoire de la Côte d'Azur, CNRS, Laboratoire Lagrange, Bd de l'Observatoire, CS 34229, F-06304 Nice Cedex 4, France.

*Corresponding author(s). E-mail(s): jlrizos@iaa.es;

**Abstract**

**Jupiter's irregular satellites are thought to be relics of early Solar System planetesimals. However, their small sizes, large distances from Earth, and close angular proximity to Jupiter, make them difficult to characterize remotely. Here we report multi-instrument observations of km-sized Kallichore, the only irregular moon of Jupiter amenable to a close flyby of ESA's Jupiter Icy Moons Explorer (JUICE), whose astrometric and physical properties were still poorly constrained. We used Hubble photometry and astrometry, followed by a ground-based multi-site stellar occultation campaign and by astrometric and photometric observations from the 10.4-m Gran Telescopio de Canarias. We could reduce Kallichore's orbital uncertainty by up to ∼80% and determine Kallichore's shape: an elongated object with a minimum semi-axis ratio a/b = 1.53 ± 0.10, an area-equivalent diameter of $3.8^{+2.3}_{-0.3}$ km and a dark surface with geometric albedo $3.7\%^{+0.7}_{-2.2}\%$. No close companions were detected. These new constrains provide a viable pathway toward a JUICE flyby once the spacecraft arrives at the Jupiter system in 2031.**

**Keywords:** Jovian Irregular Satellites, JUICE Mission, Stellar Occultations

# 1 Main Text

Jupiter hosts a large population of irregular satellites, numbering 107 objects at the time of writing (28 May 2026; https://www.minorplanetcenter.net/iau/NatSats/NaturalSatellites.html). The majority of these moons cluster into four main dynamical groups: the prograde Himalia group and the retrograde Ananke, Carme, and Pasiphae families [1], although recent discoveries suggest a more complex structure, with additional subgroups and possible new families [2]. These groupings are based on similarities in their orbital parameters and are reinforced by shared physical properties such as colour, albedo, or spectra [3–9]. However, given that these objects are kilometre-sized, intrinsically faint at heliocentric distances of ∼4 AU, and further affected by scattered light from nearby Jupiter, substancial observational biases are expected. Current surveys are likely complete only down to ∼24 magnitudes in the $R$-band, corresponding to diameters of order ∼2 km for an assumed albedo of ∼0.05 [2]. This implies that a substantial population of smaller objects may remain to be discovered. Over the past two decades, photometric and spectroscopic studies have revealed a wide diversity in the colours and compositions of Jupiter's irregular moons. Some objects, such as Kalyke, exhibit spectral similarities to Centaurs and trans-Neptunian objects (TNOs) [3], whereas others show affinities with Jupiter Trojans [9]. Together, these findings suggest capture from diverse source populations in the early Solar System [10], including possible contributions from TNOs and Trojan-like reservoirs, as described by the Nice model [11] during the early phase of planetary migration. Moreover, high-quality spectra of the Himalia family provide compelling evidence for the collisional fragmentation of a differentiated progenitor, with Himalia representing the core, Elara the mantle, and Lysithea the crust [7, 9]. Consistent with this scenario, the captured bodies are expected to have been subsequently shaped by collisional fragmentation. As such, Jupiter's irregular moons may offer unique access to primordial outer Solar System material now located much closer than that of trans-Neptunian populations.

Kallichore (also known as Jupiter XLIV or S/2003 J11), discovered in 2003 (https://www.minorplanetcenter.net/mpec/K03/K03E29.html) is an irregular satellite orbiting at a distance of approximately 24 million km from Jupiter, with an estimated diameter of only ∼2 km [12, 13]—although more recent estimates suggest a larger size of 3.1–4.3 km, assuming $p_V = 0.06 - 0.03$, respectively, where the $p_V$ is the V-band geometric albedo [14]. Kallichore's orbital parameters indicate that it belongs to the Carme group [12], a swarm of small satellites thought to be fragments of the moon Carme (from which the group takes its name), which has a diameter of $46.7 \pm 0.9$ km and a geometric visible albedo of $3.5 \pm 0.6\%$ [15]. The existence of this group is supported by their shared spectral properties: all members exhibit red colours, with spectra similar to those of D-type asteroids [3, 6], although their $V - R$ indices, with $V$ and $R$ being the magnitudes in the V and R photometric bands, respectively, are not uniform, with values such as $0.47 \pm 0.02$ (Carme), $0.52 \pm 0.04$ (Taygete) and $0.72 \pm 0.09$ (Kalyke) [3].

As the sole irregular satellite accessible along the planned trajectory of ESA's Jupiter Icy Moons Explorer (*JUICE*) mission [13, 16], Kallichore is of particular interest and a flyby is being evaluated [13, 14, 17]. In JUICE's reference trajectory (*CReMA_5_1*), the closest approach is expected on 7 October 2031 at a distance of nearly one million kilometres [14]. However, resolved imaging with JANUS and MAJIS, instruments on board *JUICE*, requires flyby distances below $10^5$ km and $10^4$ km, respectively [14], motivating ongoing assessments of spacecraft trajectory adjustments toward a flyby at $\sim 10^3$ km. In such a retargeted scenario, the encounter would instead occur on 5 October 2031, with a closest-approach distance of 700 km [13]. The trajectory is currently planned to be frozen by the end of 2028 [18], although fine-tuning may still be possible in the months prior to the flyby, based on updated target ephemerides and JUICE navigation.

The opportunity to study this object in situ by *JUICE* would provide invaluable insights into its surface composition through spectral measurements, allowing the identification of key minerals (e.g., silicates, pyroxenes, olivine), the detection of possible organic compounds or hydrated minerals, and even the potential detection and characterisation of water ice within permanently shadowed regions (cold traps). In addition to compositional analysis, high-resolution imaging can reveal signatures of space-weathering processes, helping to assess environmental effects at this heliocentric distance and providing valuable context for the results of the *JUICE* mission to the Galilean moons and the Lucy mission to Jupiter's Trojan asteroids. The flyby could also provide constraints on the cratering age of the surface, and allow for shape reconstruction and the study of blocks and boulder distributions, all of which would contribute to understanding the object's geological and collisional history. *JUICE*'s remote sensing instruments would allow high-resolution ($\sim$10 m/pixel) multispectral imaging in the VIS-NIR range, spectral mapping from the VIS to NIR, and UV and microwave spectra of the surface and surrounding environment. Depending on the flyby distance and satellite characteristics, other instruments on-board *JUICE* may provide additional data on Kallichore's subsurface structure, its gravity and its effects on the local Jovian magnetospheric environment. This would be only the second-ever close flyby of an object of this class, following the *Cassini* (NASA-ESA-ASI) spacecraft's encounter with Phoebe in 2004 [19], and would significantly advance our understanding of these elusive and small-sized populations. Despite this large difference in size (Phoebe has a diameter of approximately 220 km), the expected spatial resolution is comparable to that achieved by *Cassini* at Phoebe ($\sim$13–27 m/pixel), implying that the irregular satellite could be resolved with several hundred pixels across, placing this study in a scarcely explored observational regime.

However, observing Kallichore from the ground is highly challenging, and since its discovery in 2003 and up to the time of writing, only 50 astrometric measurements have been reported to the Minor Planet Center (MPC), with the most recent obtained in September 2021 (https://data.minorplanetcenter.net/explorer/?tab=Designated&search=Jupiter+44). This, based on the orbital fit described in Methods, propagates into a positional uncertainty (1-$\sigma$) of 1544 km during the potential close approach in 2031. *JUICE* ideally requires positional uncertainties below $\sim$30 km to ensure that the target lies within the instruments' field of view during the flyby, although there is some flexibility in the observation strategy, albeit at the cost of a reduced effective spatial resolution.

To support the *JUICE* mission in this effort, we employ the stellar occultation technique, which measures the flux drop of a background star as it is occulted by a small body. The sub-milliarcsecond astrometric precision provided by the *Gaia* catalogue (DR3; [20]) can be directly transferred to the occulting object, while simultaneously constraining its size, shape, and albedo. This technique has already played a key role in supporting other space missions. A notable example is (486958) Arrokoth, a $\sim$36 km object for which occultation measurements in 2017 [21] were critical to the successful *New Horizons* flyby in 2019. The same technique supports the Emirates Mission to the Asteroid Belt through the prior characterisation of the ∼50 km-radius asteroid (269) Justitia [22], as well as for the flyby of the *Lucy* mission to the Jupiter Trojan (15094) Polymele, an object with an equivalent diameter of ∼21 km [23–26]. In addition to the high astrometric precision achieved, stellar occultation campaigns enabled sensitive searches for small satellites, rings, and dust, thereby reducing potential hazards to spacecraft.

For this purpose, we first used the SORA software [27] to search for favourable stellar occultations by Kallichore involving *Gaia* DR3 stars, identifying a highly promising event on 15 December 2025 (see Fig. 1). This event was particularly favourable because the occulted star had a *Gaia* $G$ magnitude of 13.5 ($G^*$, the velocity-corrected apparent $G$ magnitude of the occulted star [28], was 12.8), and the predicted shadow path crossed several European countries as well as the United States, regions with extensive networks of amateur and professional observers with considerable experience in observing stellar occultations. Given the shadow velocity ($v_{\rm occ} = 10.81$ km s$^{-1}$), an occultation duration of order ∼0.2 s was expected. The event was therefore accessible to telescopes with apertures smaller than 500 mm, making it well suited to observations with mobile facilities.

However, Kallichore's astrometric uncertainty on that date (314 mas, based on the orbital fit described in Methods) translated into a shadow-path uncertainty on the order of thousands of kilometers, making it unfeasible to organise an observational campaign for an object with an expected shadow width of only a few km. To overcome this limitation, we obtained astrometric observations with the *Hubble Space Telescope* (*HST*) in the days preceding the event, substantially refining and reducing the predicted path uncertainty (see Section 2.1 in Methods). Our analysis refined the occultation prediction, reducing the cross-track 1$\sigma$ uncertainty of the shadow path to 24 km in total width, i.e., ±12 km around the nominal centre line.

Although this represented a significant refinement, it still posed a substantial challenge. By comparison, the occultation of (486958) Arrokoth targeted a ∼36 km object with a cross-track uncertainty of ±44 km [21]. In contrast, the case of (269) Justitia involved a larger body of ∼50 km, for which the cross-track uncertainty was only ±3.7 km [22]. Because the width of the projected shadow is approximately equal to the size of the occulting body, the kilometre-scale case of Kallichore, combined with a shadow-path uncertainty more than an order of magnitude larger than the object itself, pushes stellar occultation techniques to their practical limits.

We subsequently organised an astronomical campaign, inviting both amateur and professional observers from the scientific community to participate by registering on the Occultation Portal [29]. This portal played a key role by centralising communications and enabling efficient data collection for subsequent analysis.

## 1.1 Results

In total, 22 observing teams participated from Türkiye, Serbia, Croatia, Italy, Switzerland, France, and the United States, with some teams deploying multiple telescopes at different locations. This coordinated effort relied entirely on community-based, voluntary observations. Unfortunately, weather conditions were highly unfavourable across much of the predicted path, preventing many observations (see Fig. 1).

Ultimately, 13 light curves were obtained, three of which resulted in positive detections with significances above $3\sigma$, all recorded at the OAVdA Observatory in Nus (Saint Barthélemy), northern Italy (see Figs. 1c and 2). Information on the observing stations is provided in Table 1. The positive event was recorded approximately 18 km from the predicted shadow centre, corresponding to a deviation of about $\sim 1.5\sigma$ from the initial prediction. No secondary $3\sigma$ detections compatible with close companions were detected. This was expected, since sub-satellite systems are believed to be unstable for kilometre-scale bodies like Kallichore under its specific tidal constraints [30].

These results allowed us to determine that the central instant of the occultation as observed from the OAVdA observatory occurred at 03:07:$37.3^{+0.2}_{-0.2}$ UTC, enabling an astrometric determination of Kallichore's position with sub-milliarcsecond precision, corresponding to a positional accuracy of order one kilometre at the distance of the object (see Fig. 3 and Section 2.2 in Methods). The occultation geometry further constrains a minimum projected size for Kallichore, yielding an area-equivalent diameter of $3.8^{+2.3}_{-0.3}$ km. Photometric analysis of the *HST* data indicates that the object has an elongated shape. We infer a semi-axis ratio of $a/b = 1.53 \pm 0.10$, an apparent AB magnitude at zero phase of $22.889 \pm 0.013$, and an absolute magnitude of $16.192 \pm 0.013$, all measured in the F606W filter. When combined with the size constraint derived from the occultation, these measurements yield an estimated geometric albedo in the F606W filter of $p_{\mathrm{F606W}} = 3.7\%^{+0.7}_{-2.2}\%$.

Complementary observations obtained with the *Gran Telescopio Canarias* (*GTC*) on 21 January 2026, days after the occultation (see Section 2.3 in Methods) indicate an apparent magnitude of $R = 22.45 \pm 0.06$ . Assuming the colour index of $V - R = 0.47$ adopted from Carme—the reference object of the dynamical group to which Kallichore belongs—this corresponds to a $V$ magnitude of $22.92 \pm 0.06$.

## 1.2 Discusssion

From an astrometric perspective (see Section 2.4 in Methods for details), our analysis demonstrates that the positional uncertainty of Kallichore can be substantially reduced through the combination of targeted space-based astrometry and stellar occultation measurements. For the epoch 2031-10-05, corresponding to the potential *JUICE* flyby, the 1-$\sigma$ positional uncertainty decreases from 1544 km for MPC-only astrometry to 265 km when combining MPC, *HST*, stellar occultation, and *GTC* data, corresponding to an overall reduction of 82.8%. Fig. 4a illustrates the temporal evolution of this improvement.

To assess the contribution of each dataset, Fig. 4b shows the increase in positional uncertainty relative to the full solution (i.e., the solution including all available datasets) when each dataset is removed in turn. The corresponding numerical values at the flyby epoch are reported in Table 2. We find that the stellar occultation provides the dominant constraint, with a significantly larger impact than the *HST* observations, and a much larger impact than the GTC observations.

Kallichore is among the smallest irregular satellites characterised via stellar occultation to date. With an area-equivalent diameter of 3.8 km, at 4.19 AU it subtends only 1.26 mas on the sky. Our measurements provide direct constraints on its size, shape, and albedo, critical parameters for planning a potential *JUICE* flyby in 2031. Beyond physical characterisation, the astrometric results demonstrate that coordinated occultation campaigns can drastically reduce Kallichore's orbital uncertainty. Continued monitoring offers a pathway toward the ephemeris precision required for mission planning. Moreover, the improved astrometry derived from this work significantly reduces the uncertainty of future predictions, enabling a tighter constraint on the shadow path. This will facilitate the deployment of multiple, well-spaced chords in forthcoming events, ultimately yielding more robust constraints on the object's shape, size, and albedo. Observing opportunities will be announced and coordinated through the Occultation Portal, where interested observers can follow predictions and contribute to future campaigns.

# 2 Methods

## 2.1 *HST* observations

To refine Kallichore's orbit prior to the predicted occultation, we obtained astrometric observations with the *Hubble Space Telescope* under GO programme 18215 (https://www.stsci.edu/hst-program-info/program/?program=18215&pi=1). Two visits were executed on 30 November (HST1) and 5 December 2025 (HST2). Images were acquired with WFC3/UVIS using the F606W filter, selected for its high signal-to-noise ratio (SNR), stable and well-characterized Point Spread Function (PSF), and suitability for high-precision astrometry. A box-dither pattern was applied between exposures to improve PSF sampling and to mitigate detector artifacts.

Each visit consisted of eight 97 s subarray exposures using the UVIS2-2K2C-SUB aperture. All observations were performed in non-sidereal tracking mode, following the apparent motion of Kallichore on the sky-plane. The observing windows were selected to prevent saturation of stars brighter than magnitude 17 and to ensure that the $80'' \times 80''$ field of view contained at least three *Gaia* DR3 reference stars for absolute astrometric calibration (see the left panel of Supplementary Figure 1).

### 2.1.1 *HST* astrometry

A total of 16 frames were initially obtained (SNR>18 per individual image). A preliminary inspection revealed that Kallichore was contaminated by a cosmic-ray hit in one exposure, which was therefore discarded. The final dataset consists of 15 frames: eight acquired on 30 November 2025 and seven on 5 December 2025.

Astrometry was performed on `FLC` images, which are calibrated exposures corrected for Charge Transfer Efficiency (CTE) losses, following the methodology of [31, 32]. For each frame, we identified unsaturated *Gaia* DR3 reference stars and propagated their catalogue positions from epoch J2016.0 to the mid-exposure time, accounting for proper motions and parallax.

Because the observations were obtained in non-sidereal tracking mode following Kallichore, background stars appear slightly trailed in the images. Stellar positions were therefore measured by fitting trailed PSFs. We adopted the empirical WFC3/UVIS PSF models provided by STScI (https://www.stsci.edu/hst/instrumentation/wfc3/data-analysis/psf), selecting, for each star, the model closest to its detector position. Trailed PSFs were constructed by convolving the empirical PSF with the apparent sky-plane motion of the background stars induced by the non-sidereal tracking of Kallichore over the 97 s exposure time. These synthetic PSFs were fitted to the stellar profiles to determine their detector coordinates $(x, y)$ and corresponding WCS positions $(\alpha, \delta)$.

This procedure was applied to all reference stars in each frame, and the astrometric offset was derived by comparing the measured positions with the propagated *Gaia* catalogue coordinates.

Kallichore itself was treated as a point source, and its position was measured by fitting an untrailed PSF. The astrometric offset derived from the reference stars was then applied to the measured position of Kallichore to obtain its final astrometric solution. The resulting positions are listed in Supplementary Table 1.

The mean astrometric observed-minus-computed (O-C) residuals relative to the JPL ephemeris solution for Kallichore (544) (`jup347_merged_DE442`), computed in an HST-centric reference frame, are $\Delta(\alpha \cos \delta) = 31.1 \pm 3.4$ mas and $\Delta\delta = -17.5 \pm 3.2$ mas. These O-C residuals were used to refine the prediction of the 15 December 2025 stellar occultation, reducing the cross-track shadow-path uncertainty on the ground to $\sim$24 km ($1\sigma$) and thereby enabling the planning and execution of the associated observational campaign.

#### 2.1.2 *HST* photometry

Photometric measurements were performed using the CTE-corrected `FLC` images. A pixel area map (PAM, https://www.stsci.edu/hst/instrumentation/wfc3/data-analysis/pixel-area-maps) correction was applied to account for the geometric distortion of the WFC3 detectors, which causes the effective pixel area to vary across the field of view.

Using the astrometric solutions for Kallichore derived for the astrometry, aperture photometry was performed on each frame. The source flux was extracted within a circular aperture of radius 2.4 pixels centred on the measured position, while the background was estimated from a concentric annulus with inner and outer radii of 15 and 20 pixels, respectively. These aperture parameters were selected to maximise the SNR for all images.

The background level was estimated as the median flux within the sky annulus, and its dispersion was characterised by the standard deviation. The net source flux for each frame was computed by subtracting the background contribution, given by the median sky level multiplied by the aperture area, from the total flux measured within the source aperture. The photometric uncertainty was then calculated as

$$\sigma = \sqrt{F_{\rm net} + N_{\rm ap}\, \sigma_{\rm sky}^2}, \tag{1}$$

where $F_{\rm net}$ is the background-subtracted source flux, $N_{\rm ap}$ is the number of pixels within the source aperture, and $\sigma_{\rm sky}$ is the standard deviation of the background estimated from the sky annulus.

The background-subtracted flux measured in each frame represents the total number of electrons accumulated within the exposure. To convert this quantity into a calibrated magnitude, the net flux was first normalised by the exposure time to obtain a count rate in electrons per second. Because aperture photometry captures only a fraction of the total source flux, an aperture correction was applied using the encircled energy (EE) fractions provided for WFC3/UVIS (https://www.stsci.edu/hst/instrumentation/wfc3/data-analysis/photometric-calibration/uvis-encircled-energy#section-a80a1e74-ab47-4617-8971-0e30680d9986). For the adopted aperture radius of 2.4 pixels ($r \sim 0.096''$), the encircled energy (EE) fraction was obtained by linear interpolation of the tabulated values, yielding EE = 0.655, which was applied to correct the measured count rates.

The corrected count rate was converted into physical flux units using the inverse sensitivity keyword `PHOTFLAM` provided with the *HST* data. This quantity provides the flux density per unit wavelength, in units of $\rm erg\,cm^{-2}\,s^{-1}\,\AA^{-1}$, corresponding to a source with constant spectral flux density $F_\lambda$ that produces a count rate of $1\ \rm e^-\ s^{-1}$ in the given instrumental configuration. Multiplying the corrected count rate by `PHOTFLAM` therefore yields the monochromatic flux density $F_\lambda$.

Flux densities were subsequently converted into AB magnitudes following the standard definition of the AB magnitude system, defined as

$$m_{\rm AB} = -2.5\log_{10}(F_\nu) - 48.60, \tag{2}$$

where $F_\nu$ is expressed in units of $\rm erg\,cm^{-2}\,s^{-1}\,Hz^{-1}$.

To convert $F_\lambda$ into $F_\nu$, we make use of the pivot wavelength, `PHOTPLAM`, which represents the effective wavelength of the filter and provides a well-defined transformation. This results in a mean apparent magnitude of $m_{\rm AB} = 23.195 \pm 0.013$ in the F606W filter at an average solar phase angle of 7.6°.

Next, a phase-angle correction was applied according to

$$m_{\rm AB}(0^\circ) = m_{\rm AB} - \beta\,\alpha, \tag{3}$$

where $\alpha$ is the solar phase angle in degrees. Assuming a linear phase coefficient of $\beta = 0.04\,\rm mag\,deg^{-1}$ [33–35], this yields a mean zero-phase F606W AB magnitude of $m_{\rm AB}(0^\circ) = 22.889 \pm 0.013$. A plot of the phase-corrected magnitudes as a function of time for each *HST* orbit is shown in Supplementary Figure 2. After correcting for the heliocentric and HST-centric distances of Kallichore at the time of the observations, this corresponds to an absolute magnitude of $H_{\rm AB} = 16.192 \pm 0.013$.

The *HST* observations span only ∼1.25 h, and this time baseline is insufficient to robustly constrain the rotation period to determine the rotational phase at the epoch of the observations. Consequently, it is not possible to establish whether the derived photometric values represent mean, upper, or lower limits.

Finally, we computed the peak-to-peak magnitude variation from the phase-corrected photometry, obtaining a minimum light-curve amplitude of $\Delta m = 0.461 \pm$ 0.073 mag. Assuming that the object was observed in a pole-off geometry during the *HST* observations, this amplitude corresponds to a projected axis ratio of $a/b = 1.53 \pm 0.10$. It indicates a significantly non-spherical shape, consistent with a moderately elongated body, as expected for a kilometre-scale object shaped by collisional processes.

## 2.2 Occultation data analysis

CCD/CMOS image sequences were calibrated for bias and flat-field using standard procedures with the AstroImageJ (AIJ) software [36], when such calibration frames were available. For observations recorded in video format, we used the Planetary Imaging PreProcessor (PIPP, https://pipp.software.informer.com) software to convert the video files to FITS format before proceeding with photometric analysis. Time-series multi-aperture differential photometry was then performed in AIJ by measuring the flux of the occulted star (blended with Kallichore) and dividing it by the flux of selected comparison stars within the field of view. This procedure minimizes systematic photometric errors arising from atmospheric transparency variations. Because the event duration was shorter than one second, any flux modulation due to the rotational variability of Kallichore is negligible and does not affect the resulting light curves. Aperture sizes and inner and outer sky-background annuli were optimized to minimize photometric noise outside the flux drop produced by the occultation. Photometric uncertainties were computed using AIJ following its standard noise model [36]. Fig. 2 shows the three normalized light curves corresponding to the positive occultation detections obtained at the OAVdA Observatory. Because the exposure times (1 and 2 s) are significantly longer than the expected occultation duration (∼0.2 s), the flux drops do not reach 100% depth, appearing instead as partial decreases in the integrated signal. The remaining datasets showed no significant flux drops within the sampled region.

To determine the radius and sky-plane position of Kallichore from the occultation data, we adopt a Bayesian framework in which each individual light-curve data point is treated as an independent observable, with each point corresponding to a single integration period. We project every integration into the sky-plane by transforming the start and end times into coordinates $(x_{\text{init}}, y_{\text{init}})$ and $(x_{\text{end}}, y_{\text{end}})$. This procedure yields a sky-plane segment associated with a mid-exposure time $t_i$, a relative flux $f_i$, and an associated uncertainty $\sigma_i$.

The maximum along-track separation between the three telescopes is ∼12 m, which, given a shadow velocity of $v_{\text{occ}} = 10.81$ km s$^{-1}$, corresponds to a temporal offset of only ∼1.1 ms, comparable to the timing accuracy of the observations. The cross-track separation is ∼13 m. Given that the combination of the shadow velocity with our shortest exposure time (1 s) yields an effective spatial resolution at the kilometre scale, these separations are negligible. Therefore, we model the occultation using a circular limb, as the three positive chords obtained are effectively co-located, providing constraints equivalent to a single positive chord and precluding more complex geometric modelling of the projected shape. The model parameters are the radius $r$ and the centre coordinates $(x_c, y_c)$ in the sky-plane. To generate synthetic light curves, we compute the fraction of each integration segment that falls within the circle, $L_{\rm in}$, relative to its total length $L$. The synthetic relative flux is therefore defined as $1 - L_{\rm in}/L$ for each integration. We do not explicitly account for the finite angular diameter of the occulted star in our geometric approach. This approximation is justified because the stellar diameter corresponds to only ∼40 m at Kallichore's distance [37], and the Fresnel scale, $F = \sqrt{\lambda\Delta/2}$, where $\lambda$ is the observing wavelength and $\Delta$ is the geocentric distance to the occulting body, is at the few-hundred-metre level. In addition, as noted above, the combination of the shadow velocity and our shortest exposure time yields an effective spatial resolution at the kilometre scale. A summary of the occulted star's properties is provided in Supplementary Table 2.

We employ Bayesian inference to estimate the model parameters. This framework is advantageous as it provides a better understanding of parameter correlations and provides a robust quantification of uncertainties [38]. The objective is to characterize the posterior probability density function (PDF) $P(\theta|D)$, where $\theta = (x_c, y_c, r)$ is the parameter vector and $D = \{f_i\}_i$ represents the measured flux values. The posterior is derived via Bayes' theorem:

$$P(\theta|D) = \frac{\mathcal{L}(D|\theta)P(\theta)}{P(D)}. \tag{4}$$

Priors were constructed assuming no a priori correlation between parameters, using uniform distributions for each. For the centre coordinates, corresponding to the astrometric O–C residuals relative to the JPL `jup347_merged_DE442` ephemeris, we set limits of $x_c \in [65, 100]$ km and $y_c \in [-85, -55]$ km based on visual inspection of the projected chords. For the radius $r$, we adopt a uniform prior between 0 and 5 km, informed by previous size estimates [12, 14].

The likelihood function is defined as [38]:

$$\mathcal{L}(D|\theta) = (2\pi)^{-N/2} \left( \prod_{i=1}^{N} \frac{1}{\sigma_i} \right) \exp\left[ -\sum_{i=1}^{N} \frac{(f_i - m(t_i|\theta))^2}{2\sigma_i^2} \right], \tag{5}$$

where $N$ is the total number of data points and $m(t_i|\theta)$ is the model-predicted flux at time $t_i$. We assume that all errors are normally distributed and uncorrelated.

To sample the posterior, we used a Markov chain Monte Carlo (MCMC) method via the Python package `emcee` v3.1.6 [39], which implements an affine-invariant ensemble sampling algorithm [40]. We deployed 128 walkers, performing 7,500 burn-in steps (discarded) followed by a production run of 1,000 steps. We verified that the burn in phase is at least 20 times the estimated autocorrelation time to ensure convergence of the posterior distribution.

The resulting posterior distributions are shown in Supplementary Figure 3. Our parameter estimates are $(x_c, y_c) = (83.7^{+0.9}_{-1.2}, -67.5^{+2.5}_{-2.1})$ km and a radius $r = 1.92^{+1.14}_{-0.16}$ km. The values $(x_c, y_c)$ translate into angular residuals of $(\Delta(\alpha\cos\delta),\ \Delta\delta) = (27.50^{+0.29}_{-0.40},\ -22.18^{+0.82}_{-0.69})$ mas. This result provides a new astrometric constraint for Kallichore with sub-milliarcsecond precision. In Fig. 3 we plot this nominal model on the sky-plane, together with the observed occultation chords. We also plot random draws from the posterior sample to represent the uncertainties in our parameter estimation.

For $(x_c, y_c)$, which exhibit moderately symmetric marginalized posteriors, we report the median of the sample as the nominal value. For the radius $r$, which shows a skewed marginalized distribution, the mode is reported as the nominal value. In all cases, uncertainties correspond to the 68% highest density interval, i.e. the narrowest interval containing 68% of the posterior mass.

Supplementary Figure 3 reveals a significant degeneracy between $r$ and $y_c$; the distribution is bimodal for radii $\gtrsim 2$ km. This represents the symmetric North and South solutions inherent to single-positive chord observations. However, the degeneracy is broken for $r \gtrsim 3.6$ km, since the negative chords (observations showing no occultation features) from Tauxigny and Nederland rule out the possibility of a North solution if Kallichore is this large (see Fig. 3). The reported single $y_c$ value has asymmetric uncertainties that encompass all these scenarios. Note that although the RECON46 station was located closer to the positive chords, its SNR is too low to further constrain this result.

Furthermore, inference of $r$ establishes a robust lower limit on the projected size of Kallichore at the time of occultation. Our analysis indicates that the probability of the radius being less than 1.8 km is less than 5%. The upper bound of the distribution is constrained primarily by our prior rather than the occultation data, as the absence of negative chords to the South allows for larger radii within the prior range.

For each posterior sample of the projected circular model, we computed the ingress and egress times as observed from OAVdA. Since we do not have positive observations from other stations, the mid-time between these two instants can be considered to be the central time of the occultation $t_c$. Our Bayesian inference yields $t_c = 03{:}07{:}37.3^{+0.2}_{-0.2}$ UTC. We then computed the geocentric coordinates of Kallichore at that central time of the occultation using the JPL ephemerides (`jup347_merged_DE442`, `DE441`) and applied the astrometric O-C residuals to obtain the corrected position:

$$\alpha_{\mathrm{corr}} = 07^{\mathrm{h}}\, 39^{\mathrm{m}}\, 46.81162^{+0.00002}_{-0.00003}$$
$$\delta_{\mathrm{corr}} = +21^{\circ}\, 22'\, 57.8300^{+0.0008}_{-0.0007}.$$

With an estimate of Kallichore's size in hand, it becomes possible to infer its geometric albedo by combining the object's absolute magnitude with the solar reference flux. We note, however, that the radius is derived for the specific epoch of the stellar occultation. In contrast, the photometric measurements correspond to an average over a $\sim$1.25 h time span and were obtained several days before the occultation. As a consequence, the derived albedo should be regarded as an estimate that will need to be refined by future observations.

According to [41], the AB magnitude of the Sun observed at 1 AU in the F606W filter is $m_{\odot,\mathrm{AB}} = -26.85$. Using the absolute magnitude of Kallichore, $H_{\mathrm{AB}} = 16.192 \pm 0.013$, we can therefore estimate the geometric albedo in F606W, $p_{\mathrm{F606W}}$, as

$$p_{\mathrm{F606W}} = \left(\frac{1}{r}\right)^2 10^{0.4\,(m_{\odot,\mathrm{AB}} - H_{\mathrm{AB}})}, \quad (6)$$

where the radius $r$ is expressed in astronomical units. We propagate uncertainties by transforming the marginalized posterior sample of $r$ (Supplementary Figure 3) into a corresponding distribution of $p_{\mathrm{F606W}}$, assigning to each radius a random $H_{\mathrm{AB}}$ drawn from $N(16.192, 0.013)$. This yields $p_{\mathrm{F606W}} = 3.7\%^{+0.7}_{-2.2}\%$, where uncertainties correspond to a 68% ($1\sigma$) credible interval. The nominal value is computed for $r = 1.92$ km and $H_{\mathrm{AB}} = 16.192$. Note that, in addition to the uncertainty introduced by combining *HST* photometry and occultation measurements obtained at different epochs, this estimate relies on the absolute magnitude, for which a linear phase coefficient of $\beta = 0.04$ mag deg$^{-1}$ was assumed owing to the lack of a direct constraint for Kallichore.

## 2.3 *GTC* observations

As follow-up to the occultation, we were awarded regular observing time at the *GTC* during the second semester of 2025 (2025B), under proposal GTC11-25B. Observations were carried out with the OSIRIS instrument [42] during the night of 21-22 January 2026 using Sloan $R$-band imaging (see the right panel of Supplementary Figure 1). The seeing varied between 0.9 and 1.4 arcsec, under clear sky conditions and dark time.

A total of 50 images centred on Kallichore were acquired, each with an exposure time of 50 s, while the target remained at elevations above 70°, resulting in a total observing time of 1.4 h. Of the 50 images obtained, two were discarded due to significantly degraded seeing. We chose that integration time in order to avoid saturation of good reference stars for the photometry and astrometry. The tracking speed of the telescope was that of Kallichore, not the sidereal one, to maximize the flux on Kallichore and avoid spreading of light. This tracking rate was 0.29 arcsec/minute with a position angle of 277 degrees. As a consequence the stars are very slightly trailed, mostly along the Right Ascension axis, but this had no impact on our analysis because we used sufficiently large apertures, 1.2 times the full width at half maximum (FWHM) of the stellar PSF. The data were reduced using the standard *GTC* pipeline (https://www.gtc.iac.es/observing/datareduction.php). The flux of Kallichore (SNR>10 per individual image) was calibrated in absolute photometry by comparison with *Gaia* DR3 field stars. We used the transformation equations from [43] to convert *Gaia* $G$ magnitudes into $R$ magnitudes, taking advantage of the colours from the *Gaia* DR3 catalog. Using a color of $V - R = 0.47$, the same as that of the Carme satellite family [3], this resulted in an apparent magnitude of $22.45 \pm 0.06$ in the $R$-band.

The images, acquired in the standard 2x2 binning mode had a plate scale of 0.25 arcsec/pixel, good enough for accurate astrometry. The astrometry was performed by matching *Gaia* DR3 stars (corrected for proper motion and parallax at the time of the observations) with stars in the field of view. From the observed plate coordinates an order 3 polynomial plate solution was obtained with typically 80 reference stars. No Differential Chromatic Refraction (DCR) corrections were made because the observations were carried out almost at the zenith so the contributions of DCR were expected to be negligible compared to the centroiding accuracy.

The coordinates of Kallichore were compared to the JPL ephemeris solution for Kallichore (544) (`jup347_merged_DE442`) and the corresponding O-C residuals were derived. The measurements are reported in Supplementary Table 3. Given the relatively large scatter in the dataset, a sigma-clipping procedure with a threshold of $1.5\sigma$ was applied to remove outliers, resulting in a final sample of 44 measurements. The mean O-C residuals and their associated uncertainties were then computed from this subset, yielding average values of $\Delta\alpha \cos\delta = 19.3 \pm 7.1$ mas and $\Delta\delta = -25.7 \pm 10.1$ mas in the sky-plane.

## 2.4 Astrometric impact on orbital uncertainty

To assess Kallichore's orbital uncertainty, we numerically integrated its orbit and fitted the dynamical model to both publicly available astrometry and the new measurements presented here. We used the Brazilian Orbital Solution for Satellites software, described in [44], which implements the equations of motion of [45]. The model is formulated in a Jupiter-centric reference frame whose axes are aligned with the International Celestial Reference System (ICRS). The perturbers included in the model are the Sun, the planets, and the Galilean satellites. The model also accounts for Jupiter's oblateness, pole precession, and the zonal harmonics $J_2$, $J_4$, and $J_6$. Only Kallichore's orbit is integrated, while the positions of the perturbers are taken directly from JPL ephemeris `DE442_merged_JUP365`.

The fitting procedure follows [46]. The partial derivatives of the acceleration with respect to the initial conditions are integrated simultaneously with the dynamical model. These derivatives are then used in a least-squares adjustment of the residuals, yielding updated initial conditions and the associated covariance matrix.

As an initial step, we fitted the model to the 50 astrometric measurements reported to the MPC at the time of writing (https://data.minorplanetcenter.net/explorer/?tab=Designated&search=Jupiter+44), spanning the period 2003–2021. Ephemeris uncertainties are propagated following [47], using the covariance matrix and the integrated partial derivatives. Not all MPC observations report individual astrometric uncertainties, which are required to compute a meaningful covariance matrix. We therefore adopted a single uncertainty of 0.5 arcsec for all MPC measurements without reported uncertainties, consistent with standard practices in orbit determination (e.g., https://www.projectpluto.com/find_orb.htm).

Finally, we refitted the model including the observations presented in this work. The resulting positional uncertainty is shown in Fig. 4a, which displays the temporal evolution of the 1-$\sigma$ ephemeris uncertainty, derived from a full three-dimensional Cartesian propagation and expressed as the norm of the position uncertainty vector. The full solution ("*All*") includes the 50 MPC observations prior to this work, together with HST1 (30 Nov), HST2 (5 Dec), the occultation (Occ.), and GTC measurements. The orbital uncertainty exhibits a quasi-periodic modulation, whereby faster motion translates timing uncertainties into larger positional uncertainties. Additional observational biases also contribute to this behaviour, since Kallichore has preferentially been observed from Earth at larger angular separations from Jupiter, where contamination from scattered light is reduced. The incorporation of the new observations partially breaks this correlation, although a residual cyclic behaviour is still preserved.

To quantify the contribution of each dataset to the orbital improvement, we performed a series of fits in which each dataset was removed in turn from the full solution ("*All*") and evaluated the resulting positional uncertainty. This approach allows the individual impact of each dataset on the orbit determination to be assessed in a consistent manner. Fig. 4b shows the relative differences with respect to the full solution for each case, and the numerical values at the flyby epoch are reported in Table 2.

### 2.5 Data availability

The *HST* observations are publicly available through the eHST archive at https://mast.stsci.edu/portal/Mashup/Clients/Mast/Portal.html. The stellar occultation data are publicly available through the Occultation Portal (https://occultationportal.org/public/chords/2644), which provides extended information, including access to individual observing stations, observers, instrumental setups, and interactive zoomable maps. The *GTC* data are also publicly available through the same Occultation Portal link.

Correspondence should be addressed to J.L.R. (jlrizos@iaa.es).

## 3 Acknowledgments

This research is based on observations made with the NASA/ESA *Hubble Space Telescope* (*HST*) obtained from the Space Telescope Science Institute, which is operated by the Association of Universities for Research in Astronomy, Inc., under NASA contract NAS 5–26555. These observations are associated with program 18215. We acknowledge the *HST* team for granting Director's Discretionary Time and for their support in the execution of the observations.

We acknowledge the European Space Agency (ESA) for their support in disseminating the call for participation in the 15 December stellar occultation through social media channels, which was instrumental in recruiting observers for this campaign.

We acknowledge Bill Gray for his assistance with the orbital fitting.

The Astronomical Observatory of the Autonomous Region of the Aosta Valley (OAVdA), located in the Northwestern Italian Alps, is managed by the Fondazione Clément Fillietroz, which is supported by the Regional Government of the Aosta Valley, the Town Municipality of Nus and the Unité des Communes valdôtaines Mont-Émilius.

J.M.C. and S.Sa. thank Dr. Andrea Bernagozzi, Prof. Don Mauro Grosso, and Prof. Don Luca Peyron for their valuable contributions and feedback during the observational phase.

The authors thank the Action Pluriannuelle Incitative (API) Pro-Am initiated and supported by Paris Observatory in the ROADIES program context.

Y.G. acknowledges FASTRONE Scientific Instruments and Technology Ltd. Co. for their support.

# 4 Funding Statement

J.L.R., J.M.G.L., Y.K., J.L.O., N.M., A.A.C., L.M.L. and P.S.S. acknowledge financial support from the Severo Ochoa grant CEX2021-001131-S funded by MCIN/AEI/10.13039/501100011033.

J.L.R. and L.M.L. acknowledge financial support from grant PID2021-126365NB-C21.

J.M.G.L. acknowledges financial support from the Spanish Ministry of Universities through the university training programme FPU2022/00492.

A.R.G.J. acknowledges support from FAPEMIG through grant APQ-02987-24.

Y.K. and P.S.S. acknowledge financial support from grant PID2022-139555NB-I00 funded by MCIN/AEI/10.13039/501100011033.

A.A.C. acknowledges support from project PID2023-153123NB-I00 funded by MCIN/AEI/10.13039/501100011033 and the FSE+.

The observation of the stellar occultation by Kallichore was partially funded by a 2025 Research and Education grant from Fondazione CRT.

P.P., A.L., E.M.E. and M.P. acknowledge support from ASI-INAF contract 2023-6-HH.0.

M.F., A.Sp. and L.Z. acknowledge support from the INAF grants "Uncovering the optical beat of the fastest magnetised neutron stars (FANS)" and "Coordinated Multiwavelength Exploration of Fast Radio Bursts (COMEFAR)", based on observations collected at the Copernico 1.82m telescope (Asiago Mount Ekar, Italy), INAF - Osservatorio Astronomico di Padova.

# 5 Author contributions

J.L.R. led the project, coordinated the overall analysis, proposals, performed data reduction and analysis, and led the writing of the manuscript. J.M.G.L. contributed to proposals, data reduction and analysis and co-wrote the manuscript. Y.K. assisted in coordinating the occultation campaign through the Occultation Portal, contributed to data reduction and analysis, and participated in writing. A.R.G.J. performed the occultation predictions, contributed to proposals, provided software support and orbital integrations, and contributed to writing. J.L.O. assisted in coordinating the occultation campaign in the United States, contributed to data reduction and analysis, and to the writing of the manuscript. N.M. contributed to the astrometric analysis, assisted with the *GTC* proposal, and reduced the GTC data. A.A.C. contributed to the proposals and to writing. P.P. contributed to the proposals and to the discussion and writing of the manuscript. L.M.L. contributed to discussion and writing and provided funding and institutional support. S.P. and B.P. contributed to the *HST* proposal and to the reduction and analysis of the *HST* data. E.C.M.P. contributed to data collection and analysis. J.M.C., S.Sa., A.C., S.G., A.S., A.L., A.So., A.V., A.B., E.M.E., J.L.D., L.Z., M.Sk., M.P., M.B., M.F., R.B., S.Sp., V.N., F.C., J.S., L.M., M.Bu., M.N.B., M.A., M.N., N.R., O.E., P.L.P., R.L., S.F., V.E., V.L., Y.G., and Y.A. carried out occultation observations and performed independent analyses of their datasets. F.P., P.H., O.W., P.S.S., J.d.L., F.L.R., and F.T.R. provided institutional support and contributed to proposal preparation and observer coordination.

All authors discussed the results and contributed to the final manuscript.

# 6 Competing interests

The authors declare no competing interests.

# 7 Figures

# 8 Tables

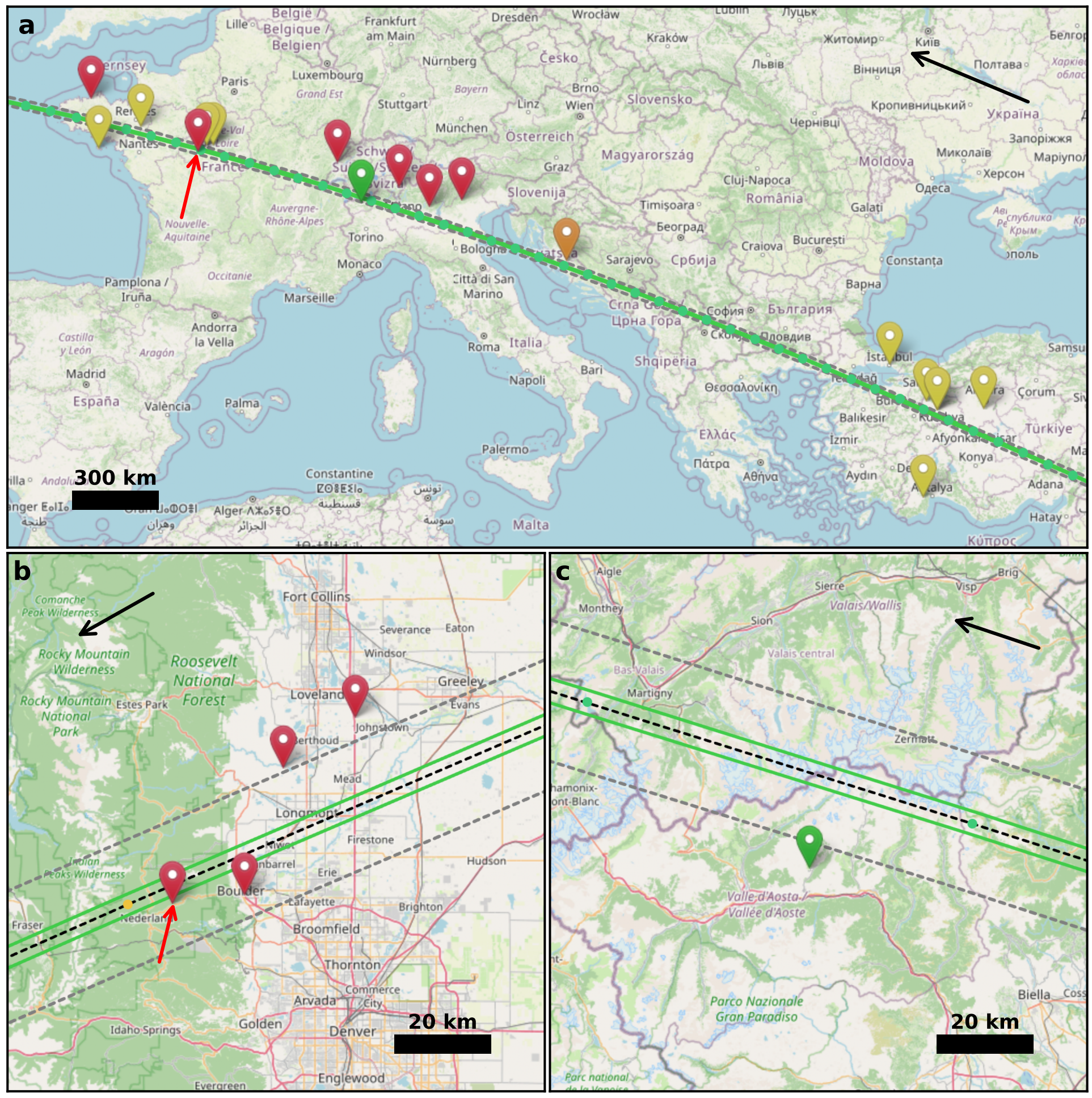

**Fig. 1**: Prediction of the stellar occultation by Kallichore on 15 December 2025. a) Shadow path across Europe, showing the location of the positive stellar occultations (green pin, OAVdA observatory). The red arrow indicates the location of Tauxigny, corresponding to the negative chord shown in Fig. 3. b) Shadow path across Colorado (United States), where four observing stations were deployed. The red arrow indicates the location of Nederland, corresponding to the negative chord shown in Fig. 3. Although the RECON46 station was located closer to the positive chords, its SNR is too low to further constrain the result. c) Detailed zoom of the region where the positive occultation detections were obtained, showing the relative location of the observatory (green pin) with respect to the predicted shadow path. Although three positive stations were observing, only one pin is visible at this zoom level due to their close proximity.
The black dashed line indicates the nominal shadow-path centre, the solid green line represents the projected shadow, adopting an area-equivalent diameter of $D = 3.8$ km for Kallichore, and the grey dashed lines show the 1-$\sigma$ uncertainty region ($\sim$24 km). Green pins denote stations with positive detections, red pins correspond to negative detections, yellow pins indicate stations affected by cloudy conditions, and orange pins represent sites impacted by technical issues. The black arrows indicates the direction of motion of the shadow, at a velocity of 10.81 $km\,s^{-1}$. Map data © OpenStreetMap contributors, licensed under the Open Database License (ODbL) v1.0. An interactive public version of this map is available through the Occultation Portal at https://occultationportal.org/public/chords/2644.

.

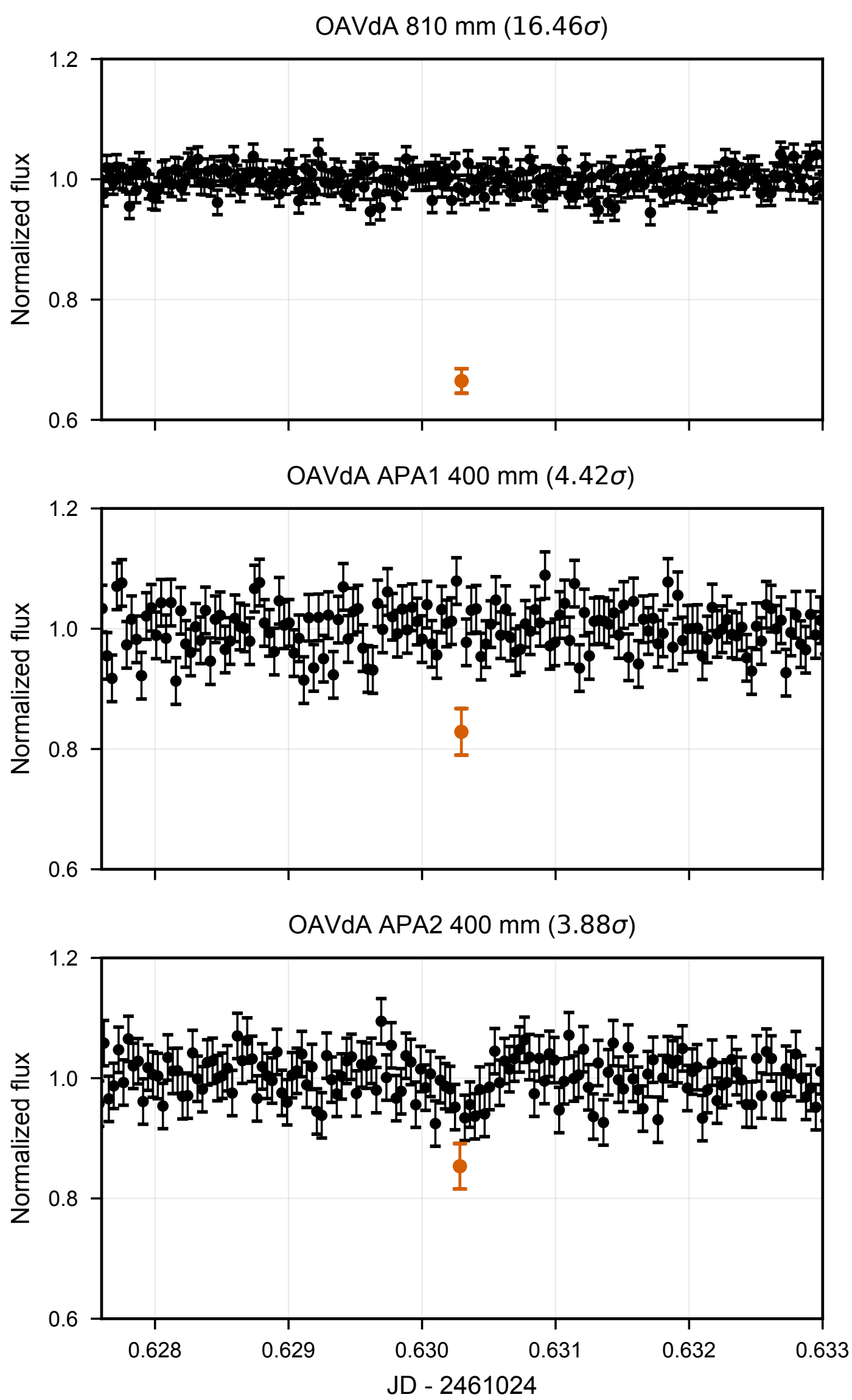

**Fig. 2**: Stellar occultation light curves from the three positive detections obtained at the OAVdA Observatory. Each panel shows the normalized flux as a function of Julian Date (JD - 2461024) for the telescopes with apertures of 810 mm (top), APA1 400 mm (middle), and APA2 400 mm (bottom). Error bars correspond to the photometric uncertainties derived from the AIJ noise model, and symbols represent the measured normalized flux values for each integration. Orange symbols highlight the data points associated with the occultation event. Above each panel, the flux drop significance (in units of $\sigma$) measured for each telescope is indicated.

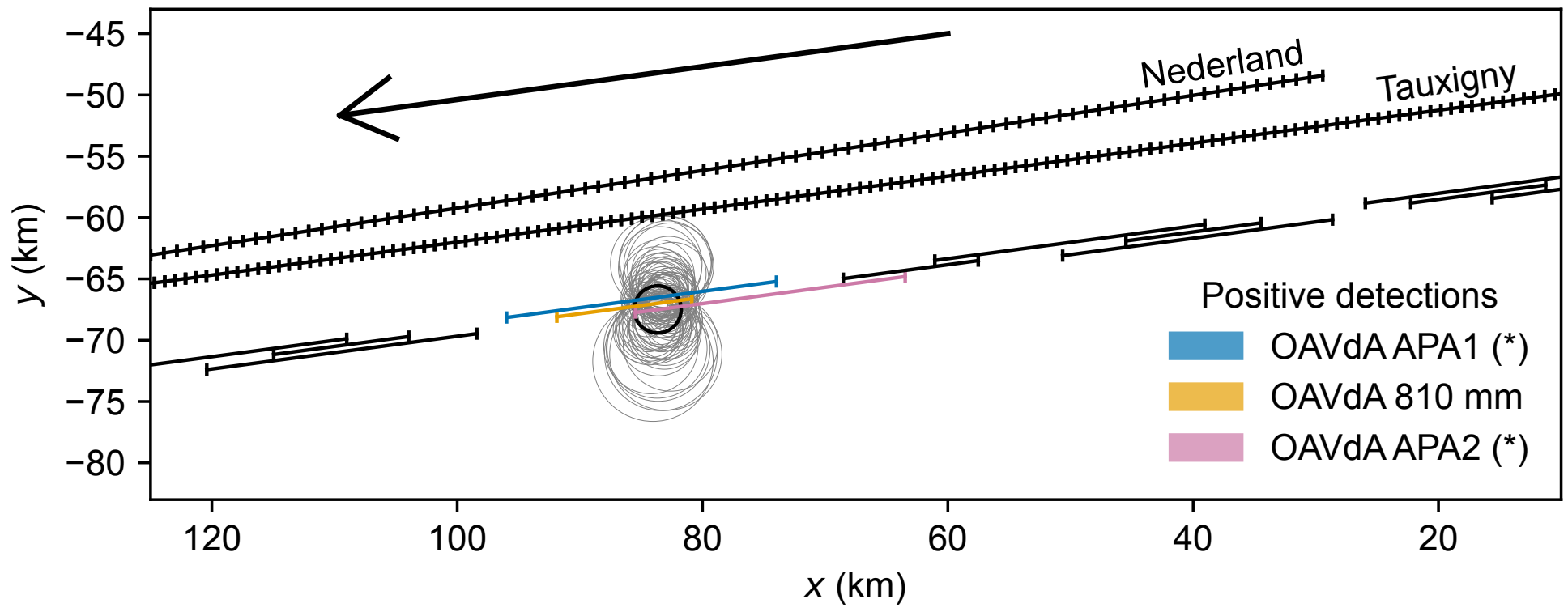

**Fig. 3**: The three positive chords and the two closest negative chords projected onto Kallichore's sky plane. The abscissa ($x$) increases eastward and the ordinate ($y$) increases northward. Each segment corresponds to a single integration (image). Colored segments indicate integrations showing flux drops above the $3\sigma$ level, corresponding the orange points in Figure 2. The thick black circle represents the nominal model reported in Section 2.2. A set of random draws from our posterior distribution is shown as thin grey circles, illustrating the associated uncertainties. As the positive chords largely overlap in the sky plane, 0.5 km vertical offsets have been applied to APA1 and APA2 in order to improve visibility, indicated in the legend with an asterisk. The black arrow indicates the direction of motion of the star relative to Kallichore.

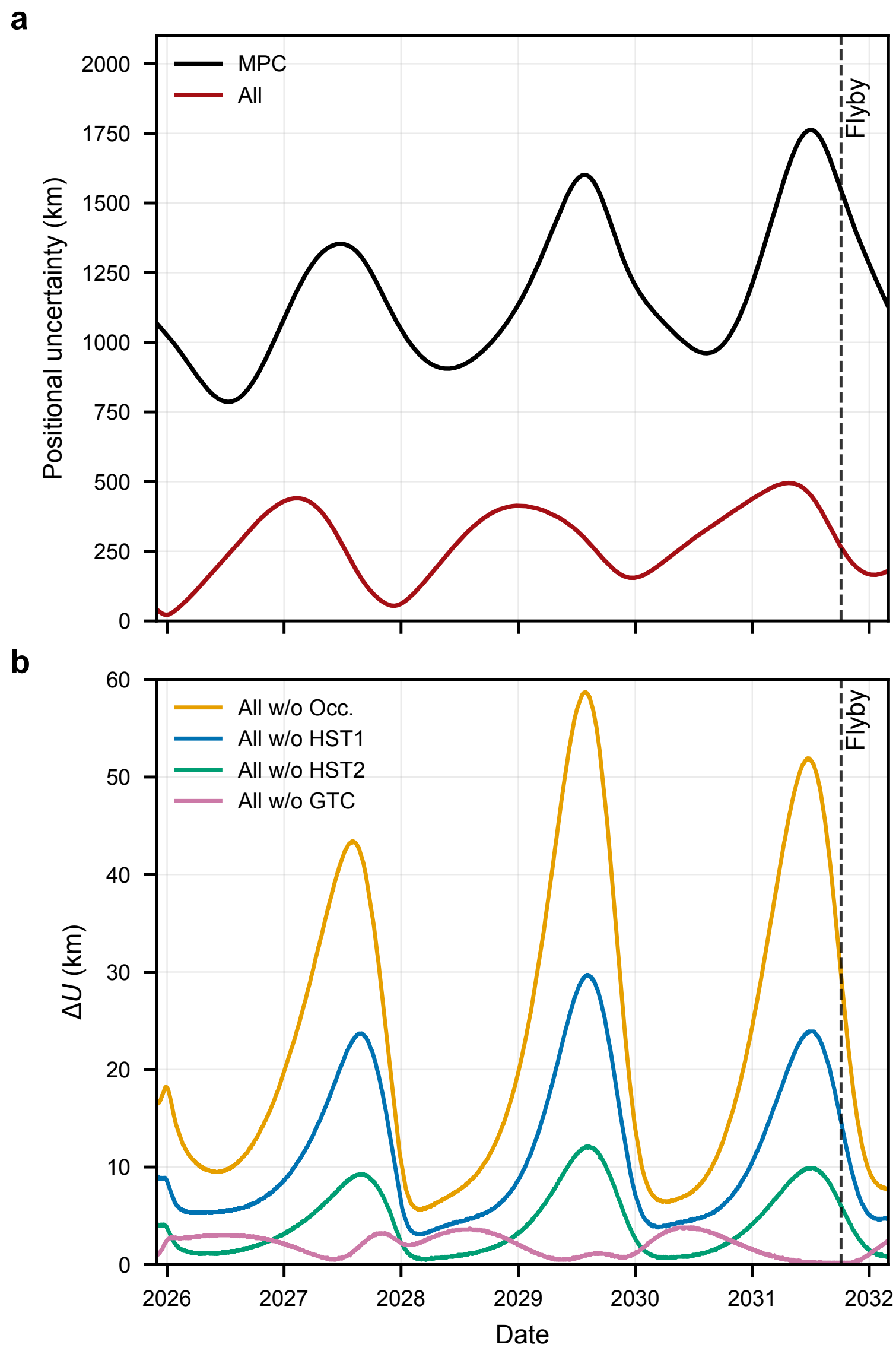

**Fig. 4**: Impact of new observations on Kallichore's orbit. (a) Evolution of the positional uncertainty of Kallichore as a function of time. The black curve shows the positional uncertainty derived from astrometric observations reported to the MPC prior to this work (2003-2021), while the dark red curve represents the updated uncertainty obtained after including our new observations (full solution). The latter, denoted in the legend as "*All*", includes MPC, HST1 (Nov. 30), HST2 (Dec. 5), occultation (Occ.), and GTC measurements. The dashed vertical line marks the epoch of the potential close flyby on 2031-10-05. (b) Relative contribution of each dataset, computed as the difference in positional uncertainty with respect to the full solution ("*All*").

**Table 1**: Observatories, locations, instrumentation, and observational configurations for the stellar occultation by Kallichore on 15 December, 2025. NTP and GPS denote Network Time Protocol and Global Positioning System, respectively. Only stations yielding positive detections or negative (non-detections) are included; observations affected by overcast conditions or technical failures are not in the table.

| Observatory | Longitude | Latitude | Altitude (m) | Aperture (mm) | Detector | Exposure time (s) | Cycle (s) | Synchronization | Result |
|---|---|---|---|---|---|---|---|---|---|
| OAVdA (Italy) | 7°28′42.671″E | 45°47′23.158″N | 1652.57 | 810 | FLI1001E | 1.0 | 2.1 | NTP | Positive |
| OAVdA (Italy) | 7°28′42.043″E | 45°47′22.648″N | 1652.57 | 400 | FLI1001E-APA1 | 2.0 | 3.2 | NTP | Positive |
| OAVdA (Italy) | 7°28′42.925″E | 45°47′22.739″N | 1652.57 | 400 | FLI1001E-APA2 | 2.0 | 3.2 | NTP | Positive |
| Lumezzane (Italy) | 10°14′22.668″E | 45°39′54.072″N | 0876.73 | 400 | QHY174GPS (CMOS) | 0.1 | 0.1 | GPS | Non-detectio |
| Asiago (Italy) | 11°34′08.397″E | 45°50′54.894″N | 1376.00 | 1820 | EFI+Aqueye+ (SPAD MPD) | 0.001 | 0.001 | GPS | Non-detectio |
| Miam-Globs (Switzerland) | 6°29′00.600″E | 46°54′36.799″N | 0975.00 | 500 | ASI (CMOS) | 0.050 | 0.050 | NTP | Non-detectio |
| Gnosca (Switzerland) | 9°01′26.616″E | 46°13′53.108″N | 0258.88 | 420 | DVTI (CMOS) | 0.040 | 0.040 | NTP | Non-detectio |
| Tauxigny (France) | 0°49′58.454″E | 47°13′24.334″N | 0091.00 | 403 | PlayerOne 533M Pro (CMOS) | 0.075 | 0.075 | GPS | Non-detectio |
| Ploumilliau (France) | 3°31′05.261″W | 48°40′09.340″N | 0096.00 | 300 | Watec 910HX-RC | 0.08 | 0.08 | NTP | Non-detectio |
| Nederland (USA) | 105°26′44.046″W | 39°59′13.955″N | 2492.62 | 400 | ASI432 (CMOS) | 0.1 | 0.1 | NTP | Non-detectio |
| Nikitin Kallichore 20251215 (USA) | 104°58′50.678″W | 40°21′03.647″N | 1536.30 | 279 | QHY174 (CMOS) | 0.066 | 0.066 | GPS | Non-detectio |
| Despil (USA) | 105°09′46.712″W | 40°15′08.425″N | 1592.10 | 280 | QHY174 (CMOS) | 0.1 | 0.1 | GPS | Non-detectio |
| RECON46 (USA) | 105°15′47.000″W | 40°00′13.000″N | 1674.40 | 280 | QHY174 (CMOS) | 0.1 | 0.1 | GPS | Non-detectio |

**Table 2**: Numerical values of the positional uncertainty of Kallichore at the flyby epoch (2031-10-05) for the different astrometric configurations shown in Fig. 4b. The full solution ("*All*") includes MPC, HST1 (Nov. 30), HST2 (Dec. 5), occultation (Occ.), and GTC measurements. Each subsequent case shows the result obtained by removing one component. The last column reports the variation in uncertainty relative to the complete set of observations ($\Delta U$), corresponding to the values shown in Fig. 4b, and enabling a direct comparison of the contribution of each dataset.

| Dataset | Uncertainty (km) | ΔU (km) |
|---|---|---|
| All | 265 | — |
| All − HST1 | 280 | 15 |
| All − HST2 | 271 | 6 |
| All − Occ. | 295 | 30 |
| All − GTC | 266 | 1 |

# Supplementary Information

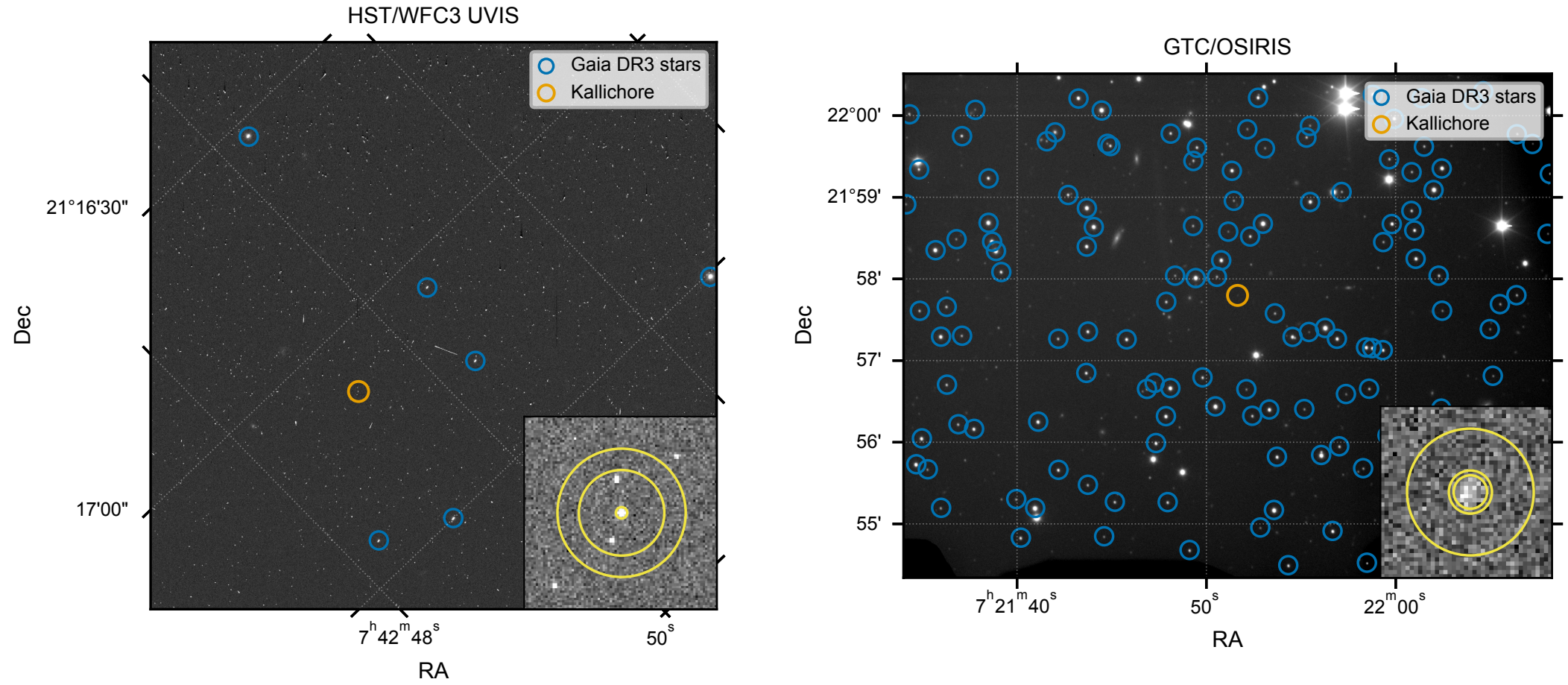

**Supplementary Figure 1**: Left: *HST*/WFC3–UVIS F606W image (ID $ifry02l1q_flc$) obtained on 5 December 2025 with an exposure time of 97 s. Right: *GTC*/OSIRIS Sloan $r$-band image (ID 0005821249) obtained on 21 January 2026 with an exposure time of 50 s. Blue circles indicate Gaia DR3 reference stars used for the astrometric solution, while the orange circle marks Kallichore. In each panel, the lower-right inset shows a zoomed view of Kallichore with the photometric apertures and sky annuli overlaid. Dotted lines represent the J2000 equatorial coordinate grid.

**Supplementary Table 1**: Astrometric positions of Kallichore measured with *HST*/WFC3.

| WFC3 dataset | UTC mid-time | $\alpha$ (h m s) | $\delta$ (° ′ ″) | $\Delta(\alpha\cos\delta)$ (mas) | $\Delta\delta$ (mas) |
|---|---|---|---|---|---|
| ifry01e5q | 2025-11-30 17:12:11.5 | 07 43 49.722290 | +21 15 22.88153 | 26.53 | -19.32 |
| ifry01e6q | 2025-11-30 17:17:32.5 | 07 43 49.654736 | +21 15 23.00223 | 26.26 | -15.42 |
| ifry01e7q | 2025-11-30 17:22:53.5 | 07 43 49.572653 | +21 15 23.18058 | 34.46 | -19.00 |
| ifry01e8q | 2025-11-30 17:28:14.5 | 07 43 49.479657 | +21 15 23.41812 | 31.60 | -14.36 |
| ifry01e9q | 2025-11-30 17:33:35.5 | 07 43 49.383380 | +21 15 23.68051 | 36.38 | -15.29 |
| ifry01eaq | 2025-11-30 17:38:56.5 | 07 43 49.289506 | +21 15 23.94798 | 33.56 | -17.12 |
| ifry01ebq | 2025-11-30 17:44:17.5 | 07 43 49.205336 | +21 15 24.19237 | 35.71 | -22.75 |
| ifry01ecq | 2025-11-30 17:49:38.5 | 07 43 49.135347 | +21 15 24.40983 | 32.53 | -13.23 |
| ifry02kuq | 2025-12-05 05:29:47.5 | 07 42 49.817452 | +21 17 06.58096 | 27.52 | -13.89 |
| ifry02kvq | 2025-12-05 05:35:08.5 | 07 42 49.740662 | +21 17 06.89942 | 31.43 | -22.77 |
| ifry02kwq | 2025-12-05 05:40:29.5 | 07 42 49.648790 | +21 17 07.30820 | 31.26 | -16.66 |
| ifry02kxq | 2025-12-05 05:45:50.5 | 07 42 49.546378 | +21 17 07.74205 | 26.60 | -22.68 |
| ifry02kyq | 2025-12-05 05:51:11.5 | 07 42 49.440400 | +21 17 08.18340 | 33.01 | -15.56 |
| ifry02l0q | 2025-12-05 06:01:53.5 | 07 42 49.239717 | +21 17 08.87009 | 31.39 | -17.70 |
| ifry02l1q | 2025-12-05 06:07:14.5 | 07 42 49.156566 | +21 17 09.06355 | 27.57 | -17.25 |

Right ascension ($\alpha$) and declination ($\delta$) are the measured equatorial coordinates in the J2000 reference frame. $\Delta(\alpha\cos\delta)$ and $\Delta\delta$ denote the astrometric O-C residuals relative to the JPL ephemeris solution for Kallichore (544) (`jup347_merged_DE442`), computed in an HST-centric reference frame.

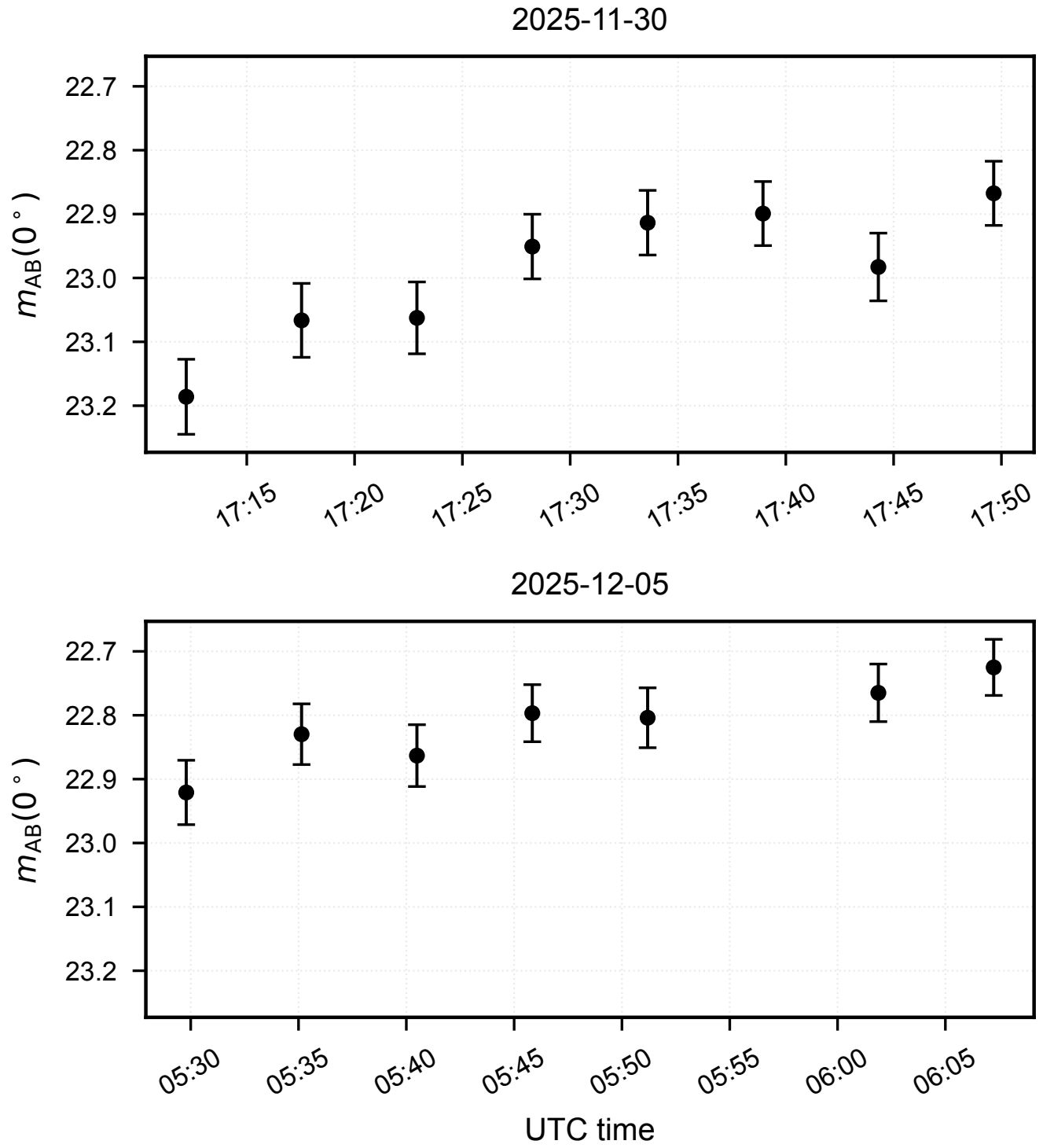

**Supplementary Figure 2**: Phase-corrected AB magnitude as a function of time for the *HST* observations. The top panel corresponds to 30 November 2025, and the bottom panel to 5 December 2025. Error bars represent the photometric uncertainties derived from the signal-to-noise ratio (SNR) photometry, and symbols correspond to the measured phase-corrected AB magnitudes for each individual exposure.

**Supplementary Table 2**: Properties of the occulted star. Coordinates are given at epoch J2016 in the ICRF, and the apparent stellar diameter is taken from [1] in the $B$ band.

| *Gaia* DR3 ID | $\alpha$ (h m s) | $\delta$ (° ′ ″) | $G, B$ (mag) | Apparent diameter (mas) |
|---|---|---|---|---|
| 673238292608047232 | 07 39 46.77065 | +21 22 56.92511 | 13.495, 15.590 | 0.0125 |

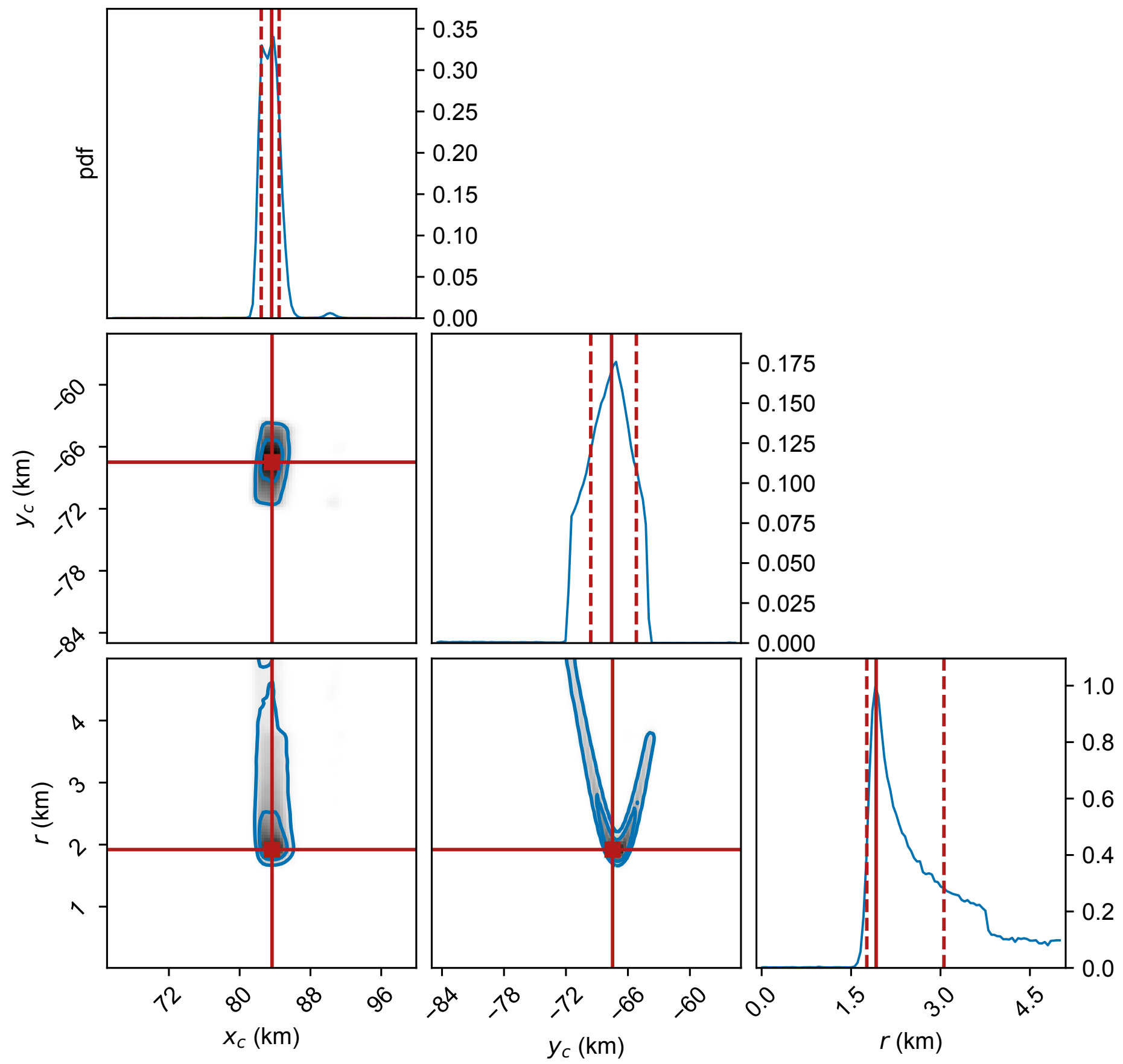

**Supplementary Figure 3**: Posterior probability distributions for the occultation model parameters. Diagonal panels show the one-dimensional marginalized posterior probability density functions for each parameter. Off-diagonal panels display the two-dimensional joint posterior distributions for all parameter pairs. The contour levels enclose the 39.3% and 86.4% credible regions, corresponding to $1\sigma$ and $2\sigma$ for a two-dimensional Gaussian distribution, respectively.

**Supplementary Table 3**: Astrometric positions of Kallichore measured with *GTC*.

| OSIRIS dataset | UTC mid-time | $\alpha$ (h m s) | $\delta$ (° ′ ″) | $\Delta(\alpha \cos\delta)$ (mas) | $\Delta\delta$ (mas) |
|---|---|---|---|---|---|
| 0005821249 | 2026-01-21 23:11:31.511 | 07 21 51.570480 | +21 57 49.96080 | 64.86 | -46.60 |
| 0005821250 | 2026-01-21 23:12:57.011 | 07 21 51.544080 | +21 57 49.94280 | 112.18 | -119.47 |
| 0005821251 | 2026-01-21 23:14:15.010 | 07 21 51.505920 | +21 57 50.19120 | -40.45 | 78.90 |
| 0005821252 | 2026-01-21 23:15:33.010 | 07 21 51.476400 | +21 57 50.13360 | -72.87 | -28.71 |
| 0005821254 | 2026-01-21 23:18:09.011 | 07 21 51.423120 | +21 57 50.46840 | -57.52 | 206.14 |
| 0005821255 | 2026-01-21 23:19:27.010 | 07 21 51.401760 | +21 57 50.34600 | 23.62 | 33.80 |
| 0005821256 | 2026-01-21 23:20:45.011 | 07 21 51.371280 | +21 57 50.36760 | -22.09 | 5.48 |
| 0005821257 | 2026-01-21 23:22:03.010 | 07 21 51.343920 | +21 57 50.28480 | -24.38 | -127.22 |
| 0005821258 | 2026-01-21 23:23:21.011 | 07 21 51.320880 | +21 57 50.29920 | 33.44 | -162.70 |
| 0005821259 | 2026-01-21 23:24:39.009 | 07 21 51.293040 | +21 57 50.73840 | 24.49 | 226.65 |
| 0005821260 | 2026-01-21 23:25:57.011 | 07 21 51.248880 | +21 57 50.46480 | -211.46 | -96.78 |
| 0005821261 | 2026-01-21 23:27:15.011 | 07 21 51.240240 | +21 57 50.70240 | 46.72 | 91.01 |
| 0005821262 | 2026-01-21 23:28:33.011 | 07 21 51.208800 | +21 57 50.61960 | -12.26 | -41.58 |
| 0005821263 | 2026-01-21 23:29:51.011 | 07 21 51.180480 | +21 57 50.78160 | -27.82 | 70.66 |
| 0005821264 | 2026-01-21 23:31:09.011 | 07 21 51.156960 | +21 57 50.63040 | 23.39 | -130.29 |
| 0005821265 | 2026-01-21 23:32:27.011 | 07 21 51.125760 | +21 57 50.72040 | -32.21 | -90.00 |
| 0005821266 | 2026-01-21 23:33:45.011 | 07 21 51.103200 | +21 57 50.90400 | 32.39 | 43.90 |
| 0005821267 | 2026-01-21 23:35:03.011 | 07 21 51.075120 | +21 57 50.81760 | 20.21 | -92.17 |
| 0005821268 | 2026-01-21 23:36:21.011 | 07 21 51.043920 | +21 57 50.97960 | -35.36 | 20.17 |
| 0005821269 | 2026-01-21 23:37:39.011 | 07 21 51.021360 | +21 57 50.90400 | 29.27 | -105.06 |
| 0005821270 | 2026-01-21 23:38:57.011 | 07 21 50.993520 | +21 57 51.04440 | 20.46 | -14.26 |
| 0005821271 | 2026-01-21 23:40:15.011 | 07 21 50.958240 | +21 57 51.15600 | -91.83 | 47.75 |
| 0005821272 | 2026-01-21 23:41:33.011 | 07 21 50.935440 | +21 57 51.13800 | -30.51 | -19.81 |
| 0005821273 | 2026-01-21 23:42:51.011 | 07 21 50.911200 | +21 57 51.24960 | 10.79 | 42.26 |
| 0005821274 | 2026-01-21 23:44:09.011 | 07 21 50.887680 | +21 57 51.20640 | 62.12 | -50.46 |
| 0005821275 | 2026-01-21 23:45:27.011 | 07 21 50.853600 | +21 57 51.35040 | -33.45 | 44.05 |
| 0005821276 | 2026-01-21 23:46:45.011 | 07 21 50.829120 | +21 57 51.39720 | 4.54 | 41.38 |
| 0005821277 | 2026-01-21 23:48:03.011 | 07 21 50.805600 | +21 57 51.35400 | 55.89 | -51.27 |
| 0005821278 | 2026-01-21 23:49:21.010 | 07 21 50.775360 | +21 57 51.54840 | 13.76 | 93.71 |
| 0005821279 | 2026-01-21 23:50:39.011 | 07 21 50.752080 | +21 57 51.45120 | 68.47 | -52.89 |
| 0005821280 | 2026-01-21 23:51:57.010 | 07 21 50.724480 | +21 57 51.54120 | 63.08 | -12.27 |
| 0005821281 | 2026-01-21 23:53:15.011 | 07 21 50.694720 | +21 57 51.54120 | 27.66 | -61.62 |
| 0005821282 | 2026-01-21 23:54:33.010 | 07 21 50.666400 | +21 57 51.58800 | 12.27 | -64.16 |
| 0005821283 | 2026-01-21 23:55:51.011 | 07 21 50.640960 | +21 57 51.71760 | 36.96 | 16.14 |
| 0005821284 | 2026-01-21 23:57:09.010 | 07 21 50.612400 | +21 57 51.75360 | 18.24 | 2.85 |
| 0005821285 | 2026-01-21 23:58:27.011 | 07 21 50.585520 | +21 57 51.71400 | 22.91 | -86.01 |
| 0005821286 | 2026-01-21 23:59:45.010 | 07 21 50.559840 | +21 57 51.80040 | 44.26 | -48.85 |
| 0005821287 | 2026-01-22 00:01:03.011 | 07 21 50.530320 | +21 57 51.85800 | 12.22 | -40.47 |
| 0005821288 | 2026-01-22 00:02:21.010 | 07 21 50.509200 | +21 57 52.00200 | 97.02 | 54.34 |
| 0005821289 | 2026-01-22 00:03:39.011 | 07 21 50.479680 | +21 57 51.98760 | 64.97 | -9.23 |
| 0005821290 | 2026-01-22 00:04:57.010 | 07 21 50.452800 | +21 57 51.99480 | 69.65 | -51.18 |
| 0005821291 | 2026-01-22 00:06:15.011 | 07 21 50.426640 | +21 57 52.04880 | 84.36 | -46.30 |
| 0005821292 | 2026-01-22 00:07:33.010 | 07 21 50.397600 | +21 57 52.09920 | 59.00 | -45.00 |
| 0005821293 | 2026-01-22 00:08:51.011 | 07 21 50.366880 | +21 57 52.01280 | 10.27 | -180.48 |
| 0005821294 | 2026-01-22 00:10:09.010 | 07 21 50.341200 | +21 57 52.10280 | 31.66 | -139.54 |
| 0005821295 | 2026-01-22 00:11:27.010 | 07 21 50.318160 | +21 57 52.27920 | 89.77 | -12.17 |
| 0005821296 | 2026-01-22 00:12:45.010 | 07 21 50.278320 | +21 57 52.39800 | -85.82 | 57.62 |
| 0005821297 | 2026-01-22 00:14:03.011 | 07 21 50.257440 | +21 57 52.26480 | 2.35 | -124.56 |

Right ascension ($\alpha$) and declination ($\delta$) are the measured equatorial coordinates in the J2000 reference frame. $\Delta(\alpha \cos\delta)$ and $\Delta\delta$ denote the astrometric O-C residuals relative to the JPL ephemeris solution for Kallichore (544) (`jup347_merged_DE442`), computed in the topocentric frame of the observer.